\documentclass[twocolumn,aps,prl,superscriptaddress,nofootinbib]{revtex4-1}
\usepackage[utf8]{inputenc}
\usepackage[T1]{fontenc}

\usepackage{hyperref}
\usepackage{amsmath, amssymb, amsfonts, mathrsfs}
\usepackage{xcolor,epstopdf}
\usepackage{exscale,times}
\usepackage{graphicx,color}
\usepackage{latexsym}

\usepackage{dsfont}
\usepackage{verbatim}

\usepackage{upgreek}

\begin{document}
	
\title{Large scale quantum walks by means of optical fiber cavities} 

\author{J. Boutari} \email{joelle.boutari@physics.ox.ac.uk}
\affiliation{Clarendon Laboratory, University of Oxford, Parks Road, Oxford, OX1 3PU, United Kingdom}
\author{ A. Feizpour}
\affiliation{Clarendon Laboratory, University of Oxford, Parks Road, Oxford, OX1 3PU, United Kingdom}
\author{ S. Barz}
\affiliation{Clarendon Laboratory, University of Oxford, Parks Road, Oxford, OX1 3PU, United Kingdom}
\author{ C. Di Franco}
\affiliation{QOLS, Blackett Laboratory, Imperial College London, London SW7 2BW, United Kingdom}
\author{ M. S. Kim}
\affiliation{QOLS, Blackett Laboratory, Imperial College London, London SW7 2BW, United Kingdom}
\author{ W. S. Kolthammer}
\affiliation{Clarendon Laboratory, University of Oxford, Parks Road, Oxford, OX1 3PU, United Kingdom}
\author{ I. A. Walmsley}
\affiliation{Clarendon Laboratory, University of Oxford, Parks Road, Oxford, OX1 3PU, United Kingdom}


\begin{abstract}
We demonstrate a platform for implementing quantum walks that overcomes many of the barriers associated with photonic implementations. 
We use coupled fiber-optic cavities to implement time-bin encoded walks in an integrated system. 
We show that this platform can achieve very low losses combined with high-fidelity operations, enabling an unprecedented large number of steps in a passive system, as required for scenarios with multiple walkers. Furthermore the platform is reconfigurable, enabling variation of the coin, and readily extends to multidimensional lattices. We demonstrate variation of the coin bias experimentally for three different values.  
\end{abstract}

\maketitle

\section{Introduction}

Quantum walks are the quantum counterparts to random walks. A quantum walker moves by superpositions of possible paths, resulting in a probability amplitude for being observed at a particular position~\cite{Aharonov1993, Kempe2003, Venegas-Andraca2012}. Quantum walks, both discrete-time and continuous-time, have been realized using cold atoms~\cite{Preiss2015, Fukuhara2013}, single optically trapped atoms~\cite{Karski2009}, trapped ions~\cite{Schmitz2009, Xue2009, Zaehringer2010}, and photons. Optical multiport interferometers provide an attractive implementation for quantum walks, since optical fields naturally exhibit coherence between pathways and allow multi-walker scenarios using multiple single photons. Optical systems have been used to implement quantum walks using bulk optics~\cite{Bouwmeester1999a, Cardano2015}, photonic chips~\cite{Perets2008, Peruzzo2010, Sansoni2012, Crespi2013}, fiber optics~\cite{Regensburger2011}, and hybrid bulk-fiber optic approaches~\cite{Schreiber2010, Schreiber2011, Jeong2013}. However, this approach is susceptible to losses, which both limit the achievable scale before being overtaken by detectors and background noise, and abrogate many of the advantages implied for applications in quantum information processing~\cite{Childs2003, Childs2004, Buhrman2006, Ambainis2007, Ambainis2003, Shenvi2003, Kempe2002, Magniez2007a, Farhi2007}.  This challenge motivates the development of low-loss, modular, guided-wave interferometer networks for quantum walks.

Quantum walks with multiple interacting walkers have been shown to realize universal quantum computation~\cite{Childs2013}. Even with multiple non-interacting walkers, quantum walks are also thought to have quantum computational power, as illustrated by the boson sampling problem~\cite{Aaronson2011}. Quantum walks with a single walker can also implement quantum computation, though this requires an exponentially large graph~\cite{Lovett2010, Childs2009a}.
A key feature of quantum walks as information processors is that they do not require time-dependent feedforward control. 

In most experimental implementations, the quantum walk takes place over spatial locations arranged in a lattice. The physical size of the lattice then determines the maximum size of the walk. This limitation can be avoided by use of optical cavities, in which case the number of physical elements required is independent of the size of the walk. This approach was first formulated for frequency-encoded quantum walks~\cite{Knight2003}. From an experimental perspective, a further key advance was the development of optical cavities that implement time-encoded walks~\cite{Schreiber2010, Regensburger2011}. 
In this case, the walker's location is represented by the time at which a pulse completes a round trip of a cavity~\cite{Schreiber2010, Regensburger2011}. In practice, the achievable scale of time-encoded walks is limited by optical loss incurred over each step. For hybrid schemes, an important challenge in this respect is the transition from bulk optics to fiber delays~\cite{Nitsche2016}. An alternate approach is complete fiber-integration, which has been developed and demonstrated with the use of optical amplifiers that counteract loss~\cite{Regensburger2011}. Amplification, however, prohibits potential extensions to single-photon and multi-walker applications.

In this paper, we implement time-encoded photonic quantum walks by means of passive fiber-integrated coupled optical cavities. 
We show that this platform achieves very low losses, reconfigurability, and high-fidelity operation over a large number of steps. In particular, we demonstrate one-dimensional coined walks of up to 62 steps with output fidelities over 0.99, for experiments with three values of coin bias in the same apparatus.

\section{Methods}

\begin{figure}[htp]
	\centering
	\includegraphics[width=0.5\textwidth]{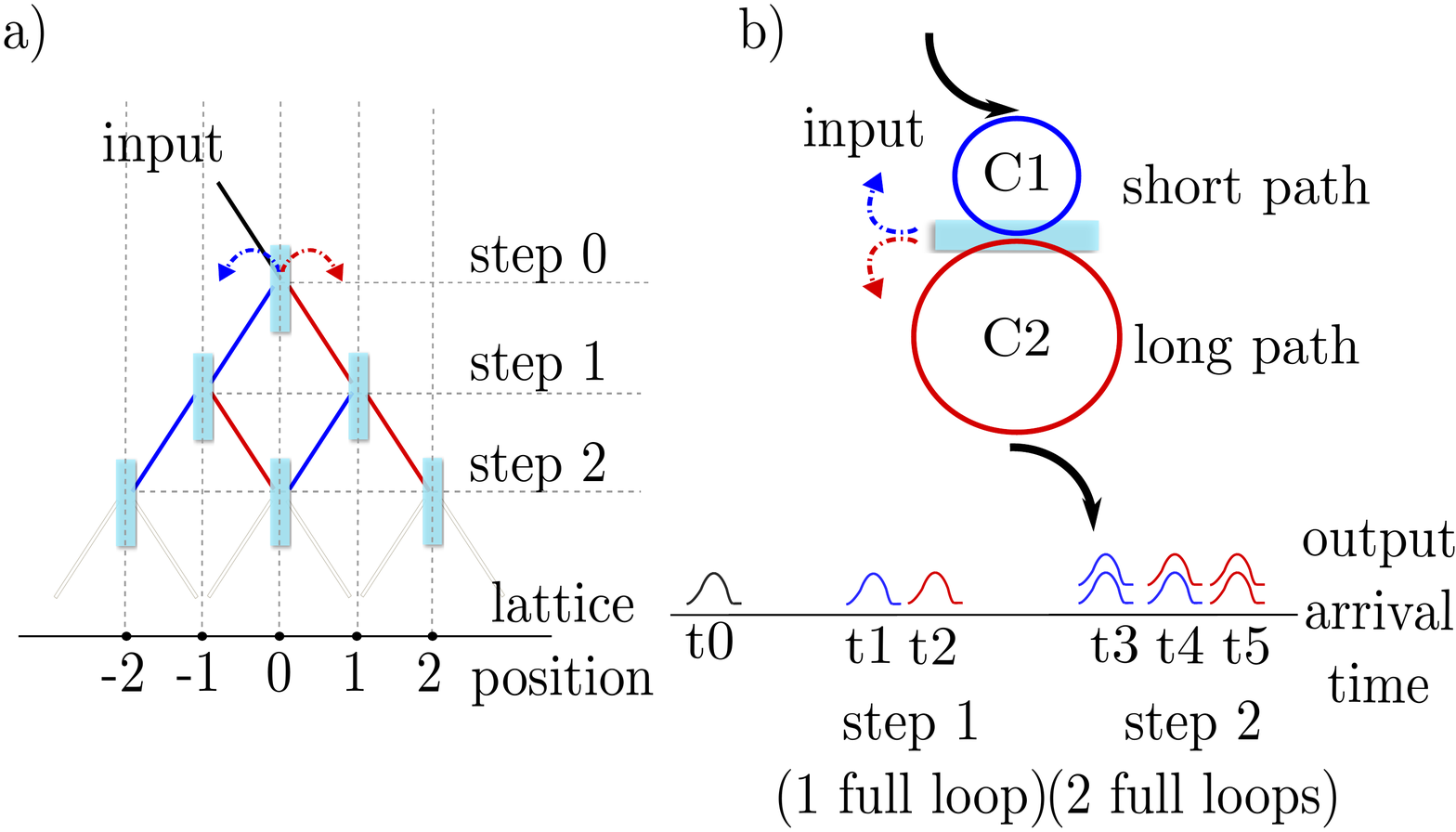}
	\caption{Optical implementations of a coined quantum walk over a one-dimensional lattice. (a) Beams propagating through a regular grid of beam splitters achieve a walk with spatial encoding. The powers of beams emerging from a grid of depth $N$ correspond to the probability distribution of the walker and coin (encoded in beam location and direction, respectively) after a walk of $N$ steps. (b) A pulse of light propagating through two coupled asymmetric ring cavities achieves a walk with time-bin and spatial encoding. The energy of pulses circulating in the cavities after $N$ round trips correspond to the walk distribution after $N$ steps. The two cavities correspond to orthogonal coin states, and timing of each pulse corresponds to the walker's position.}
	\label{Figure1}
\end{figure}

\emph{Scheme--} Our platform consists of pulsed light propagating in a network of optical ring cavities. The sizes and connectivity of the cavities can be chosen to design quantum walks with varying lattice geometry and walk rules. The simplest example, shown in Figure~\ref{Figure1},  is a coined one-dimensional quantum walk. This is achieved with two coupled cavities, which we label C$_1$ and C$_2$. These cavities have different round-trip times $T_1$ and $T_2$, which result in a round-trip delay difference  $\tau = T_2-T_1 > 0$. 
The walker is represented by an optical pulse with duration much shorter than $T_1$, $T_2$, and $\tau$.

The walk begins when a pulse of light first reaches the inter-cavity coupler that connects C$_1$ and C$_2$. The role of this coupler is to coherently divide an incident pulse into two: one remaining in the same cavity and one transferred to the other. In terms of a coined quantum walk, the coupler carries out a coin flip, for which the state of the coin (heads or tails) is encoded in the position of the pulse (C$_1$ or C$_2$). Each subsequent round trip through the cavities constitutes a new step in the walk.

Propagation of the initial pulse in the coupled cavities generates a coherent train of pulses, with pulse-to-pulse separation $\tau$. The timing of each pulse encodes the location of the walker on a one-dimensional lattice. Interference occurs when indistinguishable paths create amplitudes for the same outcome. The first interference, for example, occurs when pulses generated from the initial state traverse C$_1$ and C$_2$ each once, albeit in a different order. These two trajectories correspond to paths in which the walker moves left then right, and vice versa.

The probability distribution corresponding to measurement of the walker's position after $N$ steps is given by the relative energy of each pulse after $N$ round trips. Practically, this can be measured by coupling the cavities to an output waveguide by either an active switch or passive coupler. In the latter case, the distribution after every step can be sampled, at the cost of an effective loss that removes a fraction of the optical energy from the cavity in each step.

\begin{figure*}[t]
	\centering
	\includegraphics[width=1.0\textwidth]{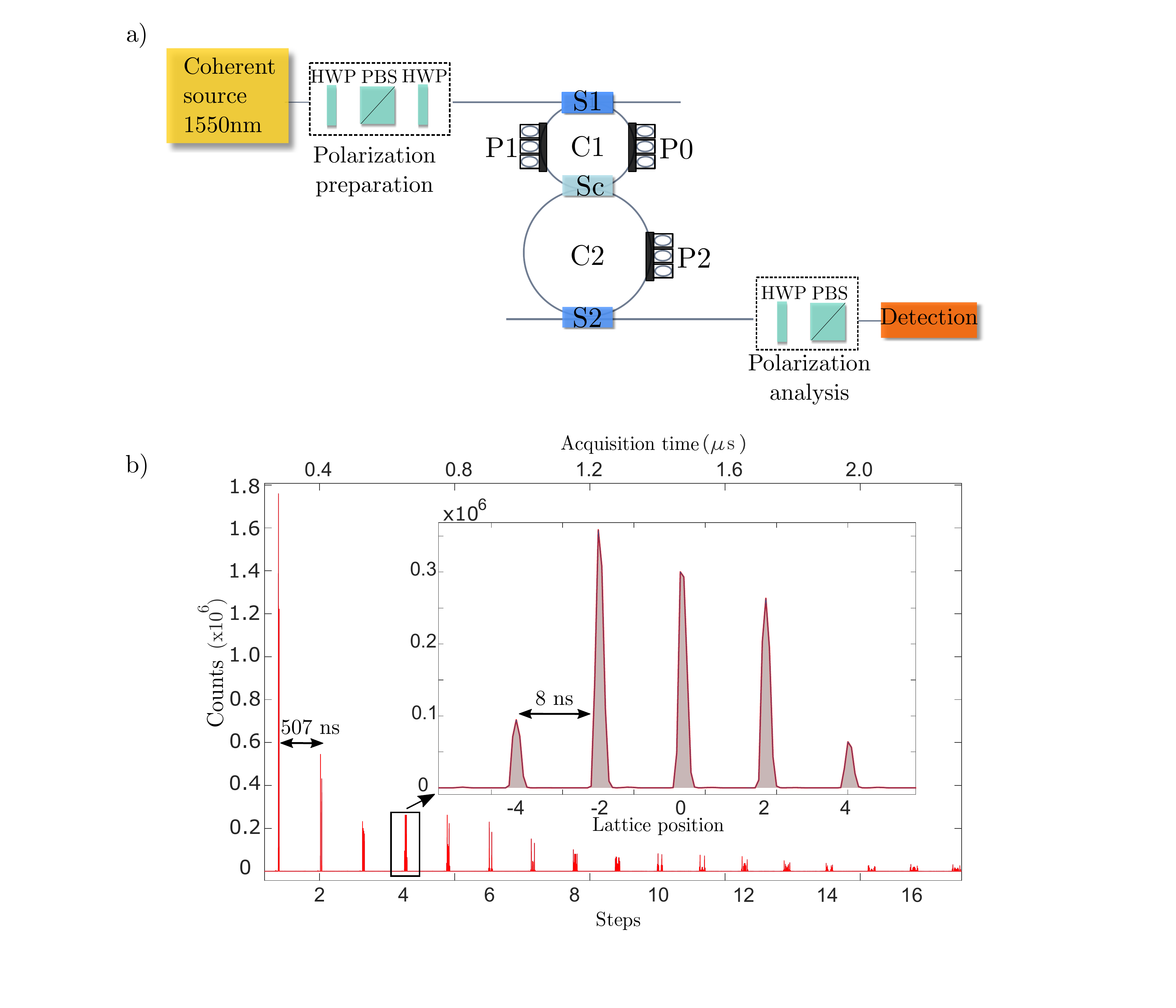}
	\caption{
		Schematic of the apparatus and representative data. (a) A 2.5-ns optical pulse is generated by a current-switched diode laser, passes through a polarization preparation stage consisting of half-wave plates (HWP) and a polarizing beam splitter (PBS), and couples into a two-cavity network via fiber coupler S$_1$. The two fiber-optic cavities C$_1$ and C$_2$ are joined by variable coupler S$_\textrm{c}$ and connected by coupler S$_2$ to an output fiber leading to a polarization analysis stage and single-photon avalanche diode. The polarization controllers P$_0$, P$_1$, and P$_2$ compensate for polarization rotations in the non-polarization maintaining fiber-optic components. (b) Histogram of the detection time from $54 \times 10^6$ trials, shown with 324\,ps bins. The inset details the fourth step. Each peak is identified with a step number and lattice position according to its detection time, as described in the main text.
	}
	\label{Figure2}
\end{figure*}

\emph{Experiment--} Our apparatus is depicted in Figure \ref{Figure2}a. First, a current-switched diode laser generates an optical pulse with a duration of 2.5\,ns and a central wavelength of 1550\,nm. A fraction of this pulse enters cavity C$_1$ through fused fiber coupler S$_1$, which has a reflectivity of 0.99. The cavities C$_1$ and C$_2$ are spliced together from standard telecommunications optical fiber (Corning SMF-28e) resulting in loop lengths (round-trip times) of 100.7\,m (503\,ns) and 102.2\,m (511\,ns), respectively. The round-trip time difference $\tau$ = 8\,ns is chosen to be larger than the pulse duration, so that the resulting time bins are distinct. The cavities are connected through an evanescent-wave fiber coupler S$_\textrm{c}$ (Evanescent Optics, model 905 SM non-PM) with a variable reflectivity $\eta_\textrm{c}$ corresponding to the coin bias.

Pulses circulating in C$_2$ are detected through the output fused fiber coupler S$_2$ with reflectivity 0.99. We choose the initial pulse energy so that the sum of all output pulses incident on the detector in one trial (i.e., all output pulses generated by one incident laser pulse) contains less than one photon on average. The pulse energies are then estimated using a single-photon avalanche diode with a time-to-digital converter that records the detection time relative to the initial pulse generation. Note that the timing window (162\,ps) and detector jitter (300\,ps) are much less than the pulse duration. Multiple trials are then used to estimate the mean photon number of each output pulse. We repeat trials every 33\,$\upmu \textrm{s}$, which limits the detection range to 62 steps in each trial. The step limit would otherwise be determined by the cavity lengths, for which the pulse trains from consecutive steps overlap after step 65.

Non-polarization-maintaining optical fiber and fiber-optic components are used throughout the apparatus to minimize optical loss. To achieve high-quality interference, however, the polarization of the pulses must be controlled. In particular, the polarization rotations experienced in each cavity must commute so that, for example, the polarization of a pulse that traverses each cavity once is independent of the order in which it does so. Polarization control is achieved with three fiber paddle polarization controllers, labelled P$_0$, P$_1$ and P$_2$ in Figure~\ref{Figure2}a. These rotations are aligned using polarization preparation and analysis stages before and after the cavities and the first three transmitted pulses. After alignment, the polarization rotation errors over each cavity is measured to be less than 0.002\,rad.

Representative data from 54$\times$10$^6$ trials are shown in Figure \ref{Figure2}a as a detection-time histogram. Time bins are observed as distinct peaks, each of which corresponds to a lattice position after a number of steps. Steps are identified as clusters of peaks that occur every 507\,ns; lattice positions are identified as peaks within those clusters with peak-to-peak separation of 8\,ns. To estimate the relative average energy of each pulse, we sum the counts in each peak, compensate for the detector reset time (which prohibits more than one detection event per trial), and subtract the average background measured between peaks. The optical loss per round trip in each cavity can then be readily determined from the exponential decay  of pulses that circulate only in that cavity-- i.e., pulses that correspond to an extremal lattice position for a given step number. We measure a round-trip excess loss, accounting for the reflectivities of the couplers, of 0.50\,dB (0.47\,dB) for C$_1$ (C$_2$). Finally, the normalized quantum walk outcome is calculated by dividing the energy of each pulse for a given step by the total energy of pulses in that step.
%
%
\section{Results}

\begin{figure*}[b]
	\centering
	\includegraphics[width=0.9\textwidth]{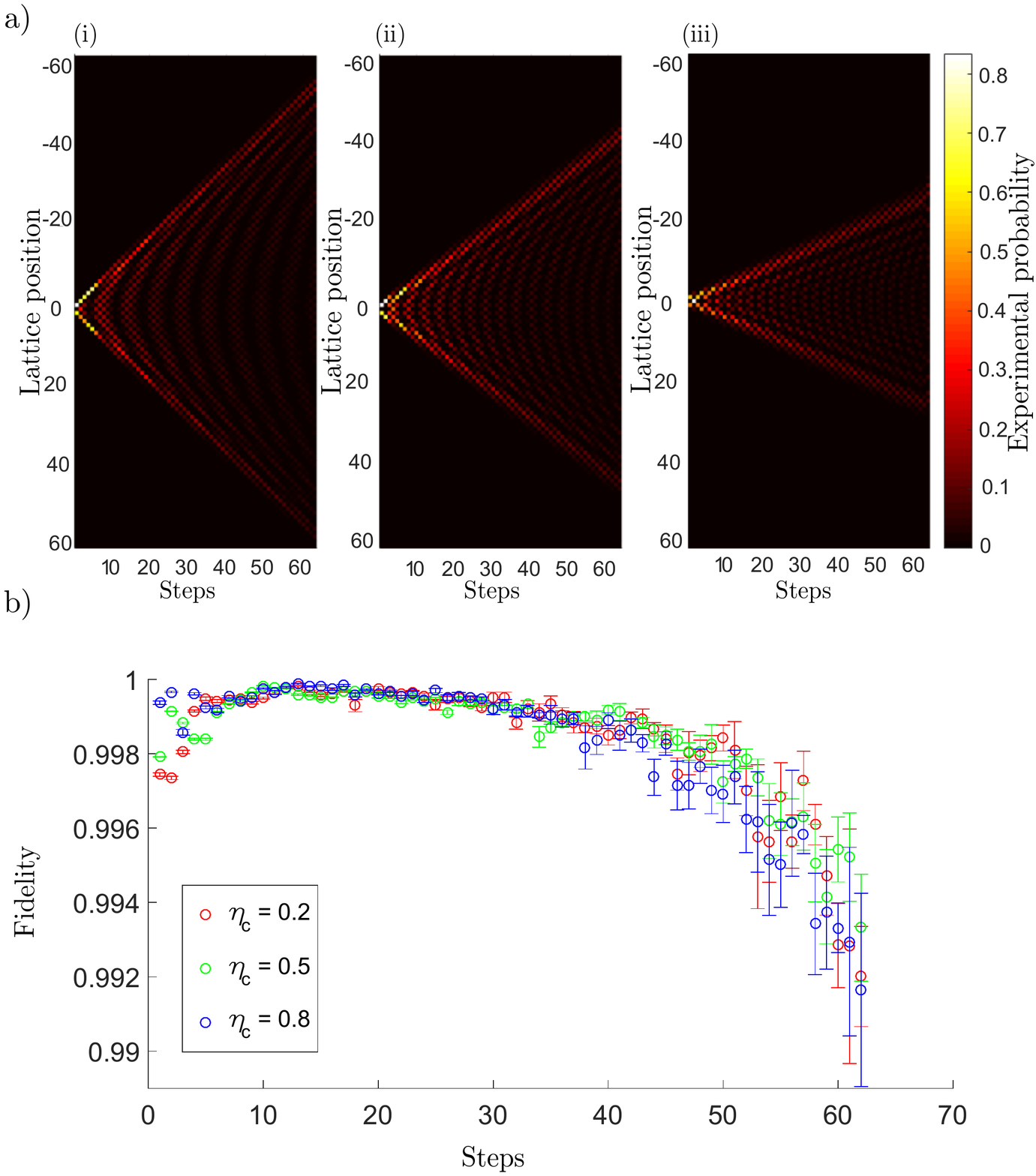}
	\caption{ Experimental results for one-dimensional coined quantum walks. (a) Measured walk outcomes for 3 different values of coin bias: (i) $\eta_{\textrm{c}} = 0.2$, (ii) $\eta_{\textrm{c}} = 0.5$, and (iii) $\eta_{\textrm{c}}= 0.8$ . (b) Fidelity of the experimental probability distribution with respect to that of an ideal quantum walk.}
	\label{Figure3}
\end{figure*}

Figure~\ref{Figure3}a shows measured quantum walk outcomes for experiments with three values of coin bias $\eta_\textrm{c}$: 0.2, 0.5, and 0.8. We obtain precise estimates of the resulting probability distribution through step 62, the maximum that can be observed in our setup with the chosen trial repetition rate. Data for step 62 include approximately 9$\times$10$^3$ detection events over 54$\times$10$^6$ trials. For this final step, the largest uncertainty in probability of an outcome is 0.004 due to optical shot noise, and the smallest probability that could be resolved above background noise is 0.001.

These measured distributions show good agreement with those of an ideal quantum walk. In Figure~\ref{Figure3}b, we show the fidelity $F = \sum_{i}{\sqrt{P_{\textrm{exp}}(i) P_{\textrm{th}}(i)}}$ of the measured $\{P_\textrm{exp}(i)\}$ and theoretical $\{P_\textrm{th}(i)\}$ distribution at each step of the walk, where the lattice positions are specified by the index $i$. The fidelity of two distributions is 1 if they are identical and 0 if they are non-overlapping. We calculate $F > 0.99$ for all observed steps in all experiments.
\section{Discussion and Conclusions}

We have presented a low-loss, flexible, passive fiber-optical platform for time-encoded quantum walks, and demonstrated its performance showing the longest one-dimensional walk yet observed. This platform benefits both from a cavity-based design, which allows the physical elements to be reused at each step, as well as practical advantages of entire fiber integration using high-quality components developed for telecommunications. We show that fiber polarization controllers can be used to precisely compensate for non-polarization-maintaining components, which allows for use of standard elements with minimal optical loss.

A number of technical improvements are possible for the system we demonstrate. The detector timing resolution and pulse duration can be brought below 100\,ps, which would allow a decrease in time-bin separation by a factor of 30 from that used. This gain could be used, for example, to reduce the trial duration or increase the number of observable steps. We estimate that the optical loss per round trip can be readily reduced below the measured 0.5\,dB by replacing couplers S$_1$ and S$_2$, which have excess loss of 0.3\,dB with, for example, the variable evanescent-wave coupler used for S$_\textrm{c}$, which has an excess loss below 0.1\,dB, as well as an ability to set an optimal coupling ratio.

\begin{figure}[htp]
	\centering
	\includegraphics[width=0.5\textwidth]{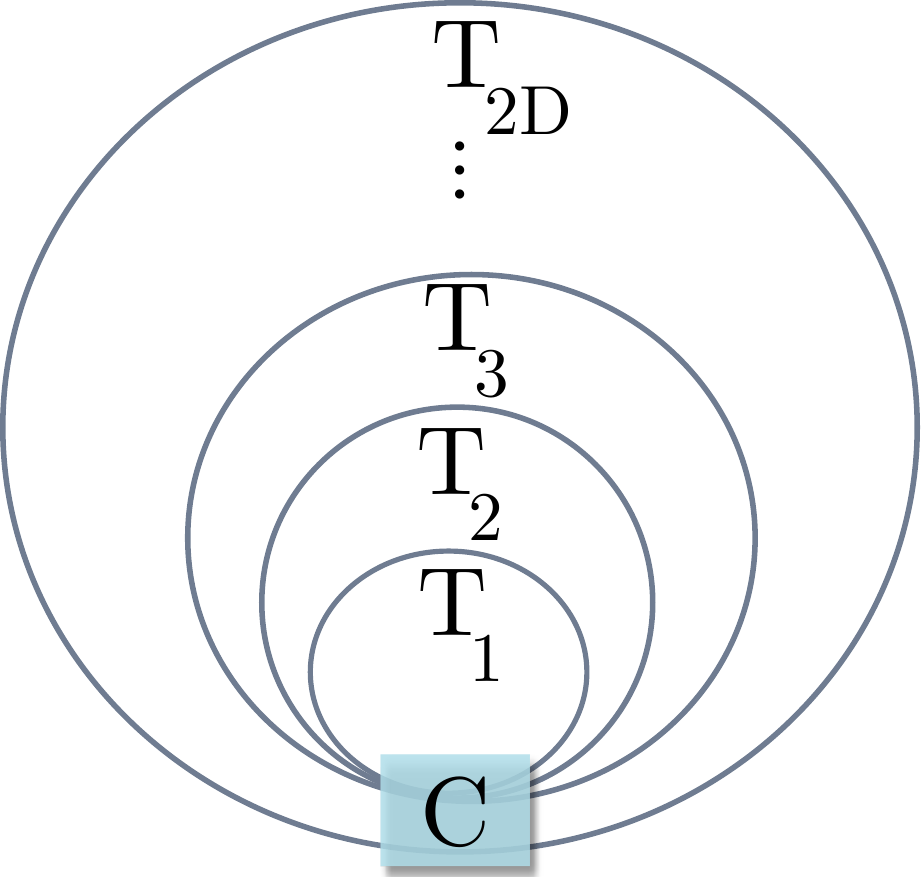}
	\caption{Schematic implementation of a quantum walk on a $D$-dimensional lattice using $2D$ ring cavities with distinct round-trip times. The coin flip is defined by the multiport coupler C connecting the cavities. Our experiment corresponds to $D=1$.}
	\label{Figure4}
\end{figure}

The use of fiber-integration with a spatially encoded coin allows our approach to be readily generalised to more complex networks of coupled cavities. Such configurations can be used to explore, for example, multi-dimensional lattices and different walk rules. For instance, a walk on a $D$-dimensional lattice can be achieved using $2D$ cavities with distinct round-trip times connected by a multiport coupler (see Figure~\ref{Figure4}). Additionally, polarization control can be included in walk design to access an additional degree of freedom. We also note that the relatively slow dynamics possible with long fiber lengths makes it possible to create a walk structure that depends on the step number and lattice position, as has been demonstrated in hybrid bulk-fiber setups~\cite{Elster2015, Nitsche2016}. A promising route for control within our all-fiber apparatus is strain-optic modulation~\cite{Humphreys2014}. The flexibility of the integrated fiber platform is in contrast to photonic chip approaches, which are limited in geometry and have not yet achieved reconfigurability in large waveguide arrays~\cite{Schreiber2011, Crespi2013}.

A future direction for this work, which we are now investigating, is studying the quantum interference of multiple single photons by connecting the input to a heralded single-photon source. This approach can enable the exploration of quantum walks with multiple walkers, and more generally, the study of time-encoded boson sampling~\cite{He2016} and related Gaussian sampling problems~\cite{RahimiKeshari2015}.

\vspace{0.5 cm}
\noindent
\section{Acknowledgements}

We thank Eilon Poem, Joshua Nunn and Henry Wen for helpful discussions.
J.B. and S.B. acknowledges support from the Marie Curie Actions within the Seventh Framework Programme for Research of the European Commission, under the Initial Training Network PICQUE (Photonic Integrated Compound Quantum Encoding, grant agreement no. 608062).
S.B. acknowledges support from the European Union's Horizon 2020 Research and Innovation 
program under Marie Sklodowska-Curie Grant Agreement No. 658073. I.A.W.  acknowledges  an  ERC  Advanced Grant (MOQUACINO) and the H2020-FETPROACT-2014 Grant QUCHIP (Quantum Simulation on a Photonic Chip; grant 
agreement no. 641039). M.S.K. acknowledge the European Research Council. I.A.W, M.S.K, and C.D.F. acknowledge the UK EPSRC project  EP/K034480/1.

\bibliography{_mybib}
\end{document}